\documentclass[10pt,pre,twocolumn,aps,amssymb,groupedaddress]{revtex4-1}
\usepackage{graphicx}
\usepackage{hyperref}
\usepackage{xr} 
\usepackage{url}
\usepackage{amsmath}
\usepackage{setspace}
\usepackage{epstopdf}
\usepackage{color}

\begin{document}
\title{Effective drag of a rod in fluid-saturated granular beds}
\author{Benjamin Allen and Arshad Kudrolli}
\affiliation{Department of Physics, Clark University, Worcester, MA 01610}

\date{\today}

\begin{abstract}
We measure the drag encountered by a vertically oriented rod moving across a sedimented granular bed immersed in a fluid under steady-state conditions. At low rod speeds, the presence of the fluid leads to a lower drag because of buoyancy, whereas a significantly higher drag is observed with increasing speeds. The drag as a function of depth is observed to decrease from being quadratic at low speeds to appearing more linear at higher speeds. By scaling the drag with the average weight of the grains acting on the rod, we obtain the effective friction $\mu_e$ encountered over six orders of magnitude 
of speeds. While a constant $\mu_e$ is found when the grain size, rod depth and fluid viscosity are varied at low speeds, a systematic increase is observed as the speed is increased.  We  analyze $\mu_e$ in terms of the inertial number $I$ and viscous number $J$ to understand the relative importance of inertia and viscous forces, respectively. For sufficiently large fluid viscosities, we find that the effect of varying the speed, depth, and viscosity can be described by the empirical function $\mu_e = \mu_o + k J^n$, where $\mu_o$ is the effective friction measured in the quasi-static limit, and $k$ and $n$ are material constants.  The drag is then analyzed in terms of the effective viscosity $\eta_e$ and found to decrease systematically as a function of $J$. We further show that $\eta_e$ as a function of $J$ is directly proportional to the fluid viscosity and the $\mu_e$ encountered by the rod.  
\end{abstract}

\maketitle

\section{Introduction}
Rod shaped solid intruders moving through granular matter immersed in a liquid can be found widely in our environment and in engineering applications ranging from the food and consumer goods industry to mechanized transportation, and biolocomotion in the sedimentary beds of water bodies. The drag acting on a rod moving through a medium is also a fundamental probe of the nature of the medium. In spite of this importance, a quantitative understanding of the drag encountered by an intruder in a granular bed immersed in a viscous fluid, and the rheology of the medium experienced by such a probe, is still lacking. 

In the case of Newtonian fluids, the drag $F_d$ experienced by a rod moving perpendicular to its axis is derived ~\cite{lamb11,kaplun57,cox70} as:
\begin{equation}
F_d = \frac{4 \pi \eta_f L U}{\frac{1}{2}-\log(\frac{L}{D})-\log(4)}\,,
\label{eq:cyldragapprox}
\end{equation}
where, $\eta_f$ is the viscosity of the fluid with density $\rho_f$, $L$ is the rod length, $D$ is the rod diameter, and  $U$ is the rod speed.
This form is considered valid for $L/D > 1$, and when the Reynolds number $Re = \frac{\rho_f U D}{\eta_f} \ll 1$ \cite{tritton59}. In the higher $Re$ inertia dominated regime, the drag becomes nonlinear and quadratic with speed. However, the rheology of granular materials immersed in a Newtonian fluid is quite different from the fluid alone, and thus the drag experienced can be quite different as well. 

The drag of a rod moving at low speeds through grains sedimented in fluids with various $\rho_f$ has been experimentally investigated~\cite{constantino11} and  found to be described by 
\begin{equation}
F_d = \mu (\rho_g - \rho_f) g D z^2,
\label{eq:schiffer}
\end{equation}
where, $\mu$ is a material dependent constant, $\rho_g$ is the density of the grains, $g$ is the gravitational acceleration, and $z$ is the penetration depth of the rod into the bed. It was reported that $\mu$ was constant at low speeds, independent of the properties of the fluid, and the effect of the fluid was to simply reduce gravity. In complementary experiments, measuring the effective friction of a flat plate moving over a fluid saturated granular bed at low speeds, it was found that the friction coefficient was the same as in the case of dry grains~\cite{siavoshi06}. The constant drag observed at low speeds in these different drag geometries show that a fluid-saturated granular medium displays a yield stress similar to that in dry granular materials~\cite{wieghardt75,vazquez10,hosoi15,faug15,maladen10,bergmann17,jslonaker17}, making it quite different from the vanishing drag experienced in a Newtonian fluid with decreasing speeds.  

In the case of dry granular beds, drag of extended objects is known to increase rapidly from the slow logarithmically increasing creep regime~\cite{reddy11}, to a more rapidly increasing drag regime with increasing speeds because of inertial effects~\cite{faug15}. Further, studies on rod drag through air moderated granular beds have reported systematic variation of drag with rod speed depending on the air speed which changes the packing and fluidization of the bed~\cite{brzinski10}.

However, the presence of the fluid can introduce rate-dependent viscous dissipation and lubrication between the grains which can further impact the encountered  drag~\cite{happel83,brady85,stevens05}.  Indeed, studies on spherical intruders moving through granular-hydrogels immersed in water found that the effective friction $\mu_e$, given by the ratio of the drag and the overburden pressure acting on the intruder, varied from being nearly constant at vanishing speeds to increasing rapidly with increasing speed~\cite{panaitescu17}. It has been also found that the granular component of the medium is essentially fluidized over the scale of the sphere diameter and the decay of the medium speed is much faster compared with a viscous Newtonian fluid~\cite{jewel18}. The granular medium used in those experiments were almost neutrally buoyant, nearly frictionless, and limited to the inertia dominated regime. Thus, systematic investigations are still necessary to measure drag over a wide range of intruder shapes and medium properties, and to identify the appropriate parameters which describe the drag experienced. 

\begin{figure*}
\includegraphics[width=16 cm]{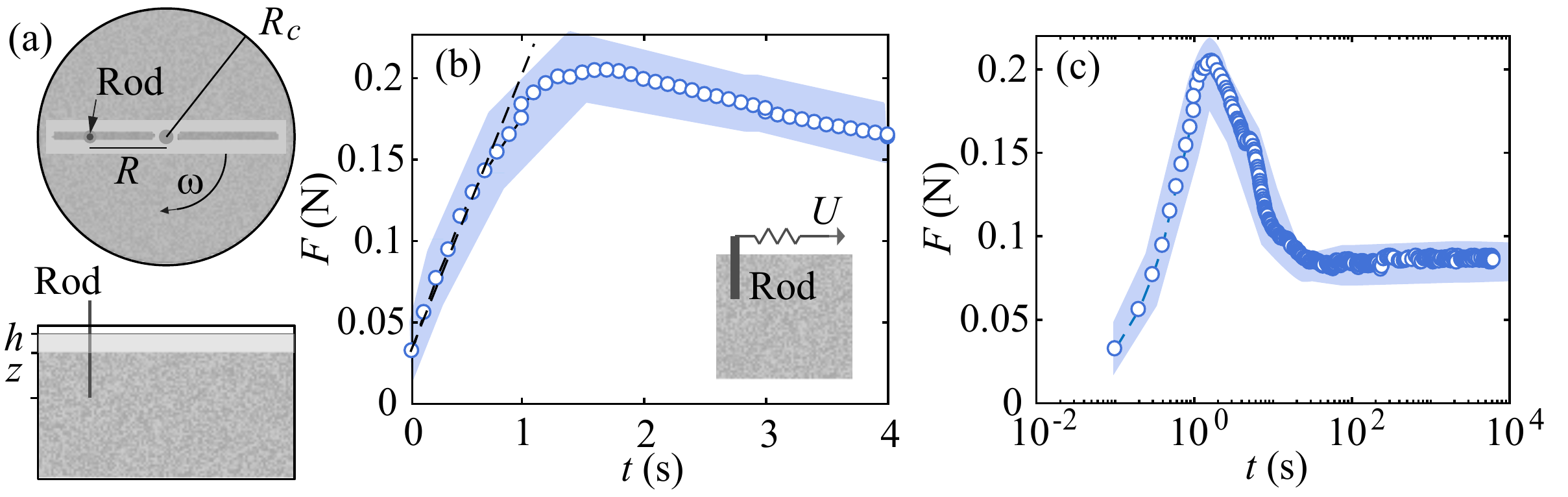}
\caption{(a) A top and side view of the experimental apparatus used to measure the drag acting on a rod inserted to a depth $z$ into a granular bed immersed in a Newtonian fluid. The apparatus is drawn to scale ($R_c = 9$\,cm). (b) The force $F$ measured as a function of time $t$ increases linearly initially, and then begins to decrease after reaching a maximum value ($U = 4.0$\,mm\,s$^{-1}$). The shaded region corresponds to the experimental error. Inset: Schematic of the rod and granular medium system of interest. (c) $F$ as function of $t$ plotted in log scale. One revolution corresponds to 100 seconds at this speed. $F$ reaches a steady value after about one revolution, and we measure the drag $F_d$ after 3 revolutions under steady state conditions.}
\label{fig:apparartus}
\end{figure*}
 
In this paper, we discuss an experimental investigation of the drag experienced by a rod in a fluid-saturated granular medium as a function of the speed of the rod, its dimensions,  and the material properties of the medium. We examine the drag experienced beyond the quasi-static regime, into the rate-dependent regime, where the drag is far greater than that required to overcome the yield stress of the medium.  We achieve this by varying the rod speed over six orders of magnitude, and the fluid viscosity over four orders of magnitude, along with the grain and rod size.  This allows us to vary the relative importance of inertia and viscosity towards identifying the appropriate non-dimensional parameters which describe the drag of the rod in fluid-saturated granular mediums. In particular, we analyze the drag in terms of an effective friction $\mu_e$ and an effective drag $\eta_e$ from the perspective of a granular medium and a viscous fluid, respectively, to understand the observed dependence with rod size $D$ and depth $z$, and fluid viscosity $\eta_f$ and grain diameter $d$.  

\section{Experimental System}
A schematic of the experimental apparatus is shown in Fig.~\ref{fig:apparartus}(a).  A granular bed consisting of spherical glass beads with diameter $d$ listed in Table~\ref{tab:grains} and density $\rho_g=2.502$\,g\,cm$^{-3}$ is filled in a cylindrical container with radius $R_{c}=9$\,cm to a height $H_c = 10 \pm 0.2$\,cm.  Besides experiments with the grains in ambient air, experiments are performed with grains immersed in various Newtonian fluids filled to a height $h=1.5$\,cm above the bed surface to avoid any capillary forces between the grains from developing. Further, $h$ is sufficiently small that the drag due to the fluid layer at the top is negligible  compared to the drag due to the granular medium. The properties of the fluids used are listed in Table~\ref{tab:fluids}.  A circular rod with diameter $D$ is then inserted to a prescribed depth $z$ into the bed. Typically grains  $d_2 \sim 150 \pm 50\, \mu$m and a rod with $D = 2.6$\,mm and $z = 3.5$\,cm are used in the discussions, unless mentioned otherwise. Thus, we are in a regime where $z \gg D \gg d$. 

The rod is moved in a periodic circular motion around the container at a distance $R$ from the center with the help of an arm attached to a stepper motor which rotates with a prescribed angular speed $\omega$ giving rise to linear rod speed $U = \omega R$. The circular nature of the system enables us to probe the drag over long times, independent of the initial preparation of the fluid saturated granular bed and initiation of motion.  The range of $R$ chosen is such that $R/D \gg 1$ and $(R_c -R)/D \ll 1$ (see Appendix~\ref{sec:side}), where the system of interest is effectively represented as a rod moving linearly through a semi-infinite bed as shown in the inset to Fig.~\ref{fig:apparartus}(b). 

\begin{table}
\begin{tabular}{cccc}
\hline
Grains & $d$ ($\mu $m) &   $d$-distribution ($\mu $m) &  $\rho_g$ (g\,cm$^{-1}$)\\
\hline
$d_1$ &  $88$ & $75-100$ & 2.502 \\
$d_2$ &  $150$ & $100-200$ & 2.502   \\
$d_3$ &  $375$ & $250-500$ &  2.502  \\
\hline
\end{tabular}
\caption{The glass beads used in the experiments.}
\label{tab:grains}
\end{table}

\begin{table}
\begin{tabular}{ccccc}
\hline
Fluid & $\rho_f$ (g\,cm$^{-3}$) & $\eta$ (mPa\,s) &  &  \\
\hline
Air & $0.0012$ & $0.018$ &  &  \\
Water & 0.998 & 1 &  &  \\
Silicone oil & 0.935 & 10 &  &  \\
Silicone oil & 0.950 & 20 &  &  \\
Silicone oil & 0.950 & 34 &  &  \\
Silicone oil & 0.965 & 100 &  &\\ 
\hline
\end{tabular}
\caption{Fluids used and their physical properties at 24$^o$C.}
\label{tab:fluids}
\end{table}

Figure~\ref{fig:apparartus}(b) shows an example of the measured force as a function of time $t$. Initially, the measured force increases as the system loads up till a maximum force is reached, after which time the force decreases slowly. Then, the measured force slowly approaches a nearly constant value as shown in Fig.~\ref{fig:apparartus}(c). This response is typical for a Coulomb frictional material which shows a higher static friction compared to a dynamic friction~\cite{nasuno98}. We focus on the steady state regime, after the initial transients have subsided, for simplicity of analysis.  In order to reduce the effect of the initial bed preparation, we move the rod thrice around in a circle before taking measurements. The drag $F_d$ is then obtained by averaging the recorded force over a 50~second time interval to average over the fluctuations which occur due to the granularity of the medium. 

\section{Drag Measurements}
\subsection{Rate Dependence}

\begin{figure}
\includegraphics[width=0.75\columnwidth]{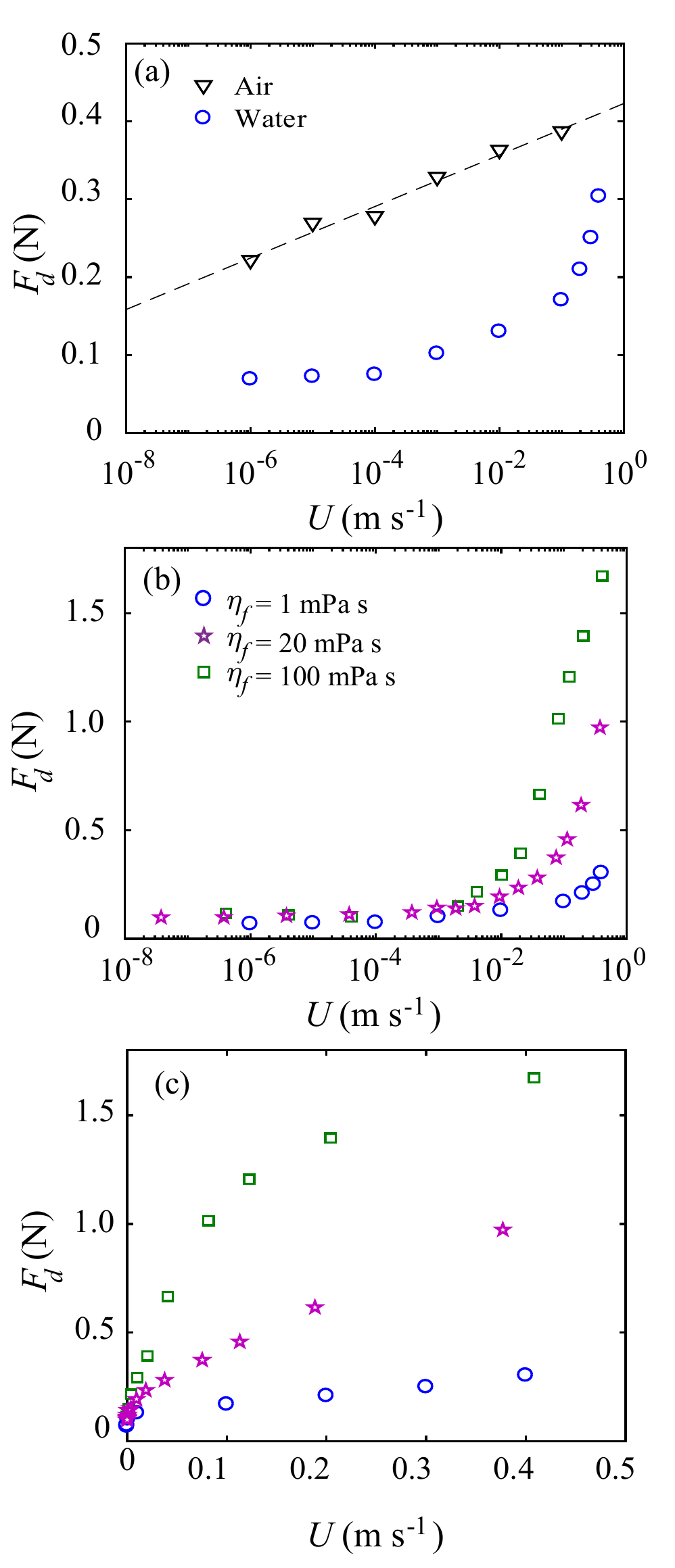}
\caption{(a) The drag of a rod measured as a function of its speed when the grains are in ambient air and when they are fully immersed in water ($z/D = 13.5$). In the case of air, the drag increases logarithmically as shown by the line fit corresponding to Eq.~\ref{eq:mue}. While drag is overall lower in case of water, a more rapid increase is observed as $U$ is increased. (b) The drag measured in fluids with similar density but different fluid viscosity $\eta_f$. (c) The same data plotted over linear scale to illustrate the sub-linear increase with speed at higher viscosities and speeds.} 
\label{fig:behavior}
\end{figure}

Figure~\ref{fig:behavior}(a) shows the measured drag $F_d$ as a function of rod speed $U$ in the granular bed when air is the interstitial fluid, and when the bed is fully immersed in water. In the case of air, we observe that the drag increases logarithmically as $U$ is increased over four orders of magnitude. {\color{black} Increase in friction have been reported in previous experiments on rod drag through a dry granular bed~\cite{reddy11}, and may be consistent with logarithmic increase in static friction observed with loading rate~\cite{rice83,heslot94,marone98,nasuno98,baumberger99}}. Thus, one may expect 
\begin{equation}
F_d = F_d (U_o) + A \log(\frac{U}{U_o}),
\label{eq:log}
\end{equation}
where, $U_o$ is a reference velocity. From the fit in  Fig.~\ref{fig:behavior}(a), we observe that the data is indeed captured by Eq.~\ref{eq:log} with $A = 0.164$, consistent with previous reports on friction~\cite{marone98,baumberger99}.   
Thus, it is possible that the slow increase in drag encountered at low speeds may be because of the solid grain-level frictional contacts in air. Whereas, the measured drag is lower in the water saturated case, and increases more rapidly as $U$ is increased.  

To understand this rate-dependence introduced by the presence of the fluid at higher speeds, we further investigate the effect of the fluid viscosity $\eta_f$ while holding the fluid density $\rho_f$ approximately constant. To highlight different aspects of the data, Fig.~\ref{fig:behavior}(b) and Fig.~\ref{fig:behavior}(c) show a plot of $F_d$ as a function of $U$ in linear-log and in linear-linear formats, respectively. We observe from Fig.~\ref{fig:behavior}(b) that $F_d$ is essentially constant and similar in value at low speeds, irrespective of the viscosity of the fluid. Whereas, $F_d$ can be observed to increase systematically faster as $\eta_f$ is increased  in Fig.~\ref{fig:behavior}(c). {\color{black} It can be also observed that $F_d$ does not increase linearly with speed, and thus the drag cannot be viewed simply as a linear superposition due to viscous and frictional contributions given by Eqs. (1) and (2), respectively.}

\begin{figure}
\includegraphics[width=0.75\columnwidth]{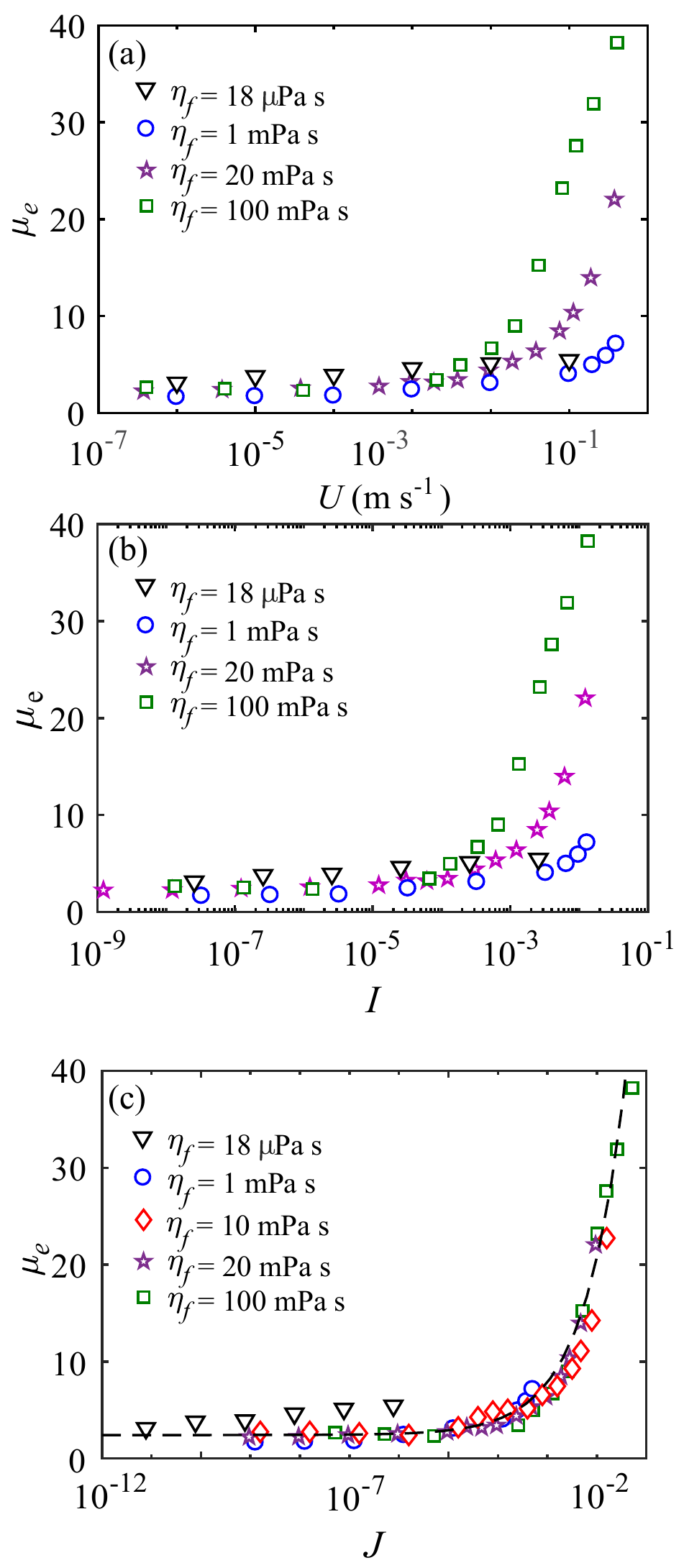}
\caption{(a) The effective friction $\mu_e$ versus $U$ is observed to converge to the same value at sufficiently low speeds, irrespective of the interstitial fluid. (b) $\mu_e$ versus inertial number $I$ varies systematically, but does not collapse due to the viscosity of the immersing fluid.  (c) $\mu_e$ versus viscous number $J$ for the various liquids is observed to collapse onto a curve for sufficiently large $\eta_f$ or low $U$. The dashed line is given by Eq.~\ref{eq:mue} with $\mu_o = 2.8 \pm 0.5,, k = 2.8 \times 10^{4}$, and $n = 0.55 \pm 0.06$. }
\label{fig:nondimensional}
\end{figure}

We use the effective friction $\mu_e$ as in previous studies~\cite{constantino11,panaitescu17} to analyze the measured drag scaled by the other relevant force in the system which corresponds to the average weight of the grains acting on the rod,     
\begin{equation}
\mu_e = \frac{F_d}{\pi \phi_g (\rho_g - \rho_f) g z^2 D/2}.
\label{eq:mue}
\end{equation}
We plot $\mu_e$ as a function of $U$ in Fig.~\ref{fig:nondimensional}(a), and observe that $\mu_e$ approaches the same value at the lowest speeds in all the  cases, irrespective of the density of the fluid. Then, we observe that $\mu_e$ increases rapidly at progressively lower $U$ as $\eta_f$ is increased. Thus, we understand the lower drag measured at the lowest speeds in Fig.~\ref{fig:behavior}(a) to be due to the reduction of the normal stress acting on the rod due to the buoyancy of the grains in the fluid,  similar to the conclusions reached by previous work~\cite{constantino11}. Otherwise, the role of the saturating fluid appears to be to increase the drag experienced by the rod when the speed is increased. 


To understand the increase of $\mu_e$, we examine the inertial and viscous time scales in the system in relation to the time scale over which the rod advances. In case of time-independent uniformly sheared dry granular materials, the inertial number was given by~\cite{dacruz05} $I = \dot{\gamma} d/\sqrt{P/\rho_g}$, where $\dot{\gamma}$ is the shear rate corresponding to a uniform shear applied between two planes, and $P$ is the normal pressure applied on the granular medium across those planes. {\color{black} However, the flow of the medium around an advancing rod is non-uniform and time-dependent.  The grains in front of the rod are accelerated from rest as they move around the advancing intruder, before slowing down and coming to rest as the intruder moves past, creating a non-homogeneous and non-steady state dynamic.} Nonetheless, this proposed definition was extended to spherical intruder dynamics by assuming that the $\dot{\gamma}$ was given by the velocity of the sphere which decayed over a grain diameter~\cite{panaitescu17}, or intruder diameter~\cite{jewel18}. The pressure $P$ was considered to correspond to the average overburden pressure due to the weight of the grains above the location of the intruder. Then, it can be noted that $I$ is complementary to the Froude number $Fr$, which has been used to characterize drag encountered in granular suspensions~\cite{graf70} and dry granular materials~\cite{faug15,takada16}. 
 
Applying the same approach to rods, assuming that the shear rate $\dot{\gamma} = U/D$, we have   
\begin{equation}
I = \frac{U d}{D\sqrt{P/\rho_g}}.
\label{eq:I}
\end{equation} 
Here, the average pressure acting on the rod over its length due to the weight of the granular medium is given by 
\begin{equation}
P = \phi_g (\rho_g - \rho_f) g z/2.
\label{eq:P}
\end{equation}
To capture the effective friction in uniformly sheared suspensions under normal pressure $P$ with shear rate $\dot{\gamma}$, a viscous number $J = {\eta_f \dot{\gamma}}/{P}$ has been proposed when viscous forces become important~\cite{boyer11}.  Then, substituting $\dot{\gamma}=U/D$, we have     
\begin{equation}
J = \frac{\eta_f U}{D P}.
\label{eq:J}
\end{equation} 

We plot $\mu_e$ versus $I$ in Fig.~\ref{fig:nondimensional}(b) and $\mu_e$ versus $J$ in Fig.~\ref{fig:nondimensional}(c). Clearly the data does not collapse with $I$ showing that viscous effects need to be included in analyzing the data. However, excellent collapse of the data is observed in Fig.~\ref{fig:nondimensional}(c), in the case of liquid-saturated granular beds over the entire range of $J$. Deviations can be observed in the case of $\mu_e$ measured with air as the interstitial fluid, as can be expected since $\eta_f$ is small and viscous forces can be expected to be negligible.  

In mixed grain-fluid systems under uniform shear, a combination of $I$ and $J$ have been proposed to describe regimes where both inertial and viscous forces may be important~\cite{trulsson12}. However, we do not observe any significant  improvement using that hybrid dimensionless number beyond the data collapse shown in Fig.~\ref{fig:nondimensional}(c) with $J$. 

Now, given $\mu_e$ goes to a constant non-zero value $\mu_o$ for vanishing $J$, and $\mu_e$ increases rapidly with at higher $J$, we use a form:
\begin{equation}
\mu_e = \mu_o + k J^n, 
\label{eq:fit}
\end{equation}
where, $k$ and $n$ are constants  to describe the data. {\color{black} This form corresponds to the addition of a granular quasi-static yield-stress $\mu_0$ and a rate dependent term as a function of $J$.  By fitting to the data,} we find $\mu_0 = 2.4 \pm 0.5$, $k=206 \pm 11 $, and $n=0.53 \pm 0.04$.  {\color{black}
The observed value of $\mu_o$ in the frictional limit is greater than the coefficient of friction around $0.5$ typically encountered for glass-on-glass or glass-on-metal motion. Such high values using similar definations of $\mu_e$ given by Eq.~\ref{eq:mue} have been reported previously~\cite{constantino11,panaitescu17}, and occur as a result of the geometry of the flow generated by the moving rod and dilatancy effects. Because the flow of the medium wraps around the advancing rod, the weight of the granular medium important to determining drag not only increases with depth but also is not normal to the direction of motion. These factors, besides the contribution of sliding friction tangential to the surface contribute to the larger value observed compared to sliding friction while considering solid-on-solid friction.  If the medium was to behave as a Newtonian fluid when well fluidized, one may expect the exponent $n$ to approach one or even exceed it, as inertial effects become important. We observe $n < 1$ which  corresponds to shear thinning because $J \propto \dot{\gamma}$, and thus the medium behaves similarly to a shear thinning fluid over the strongly rate-dependent regime accessed in our system. }

\subsection{Particle size dependence}
\begin{figure}
\includegraphics[width=0.75\columnwidth]{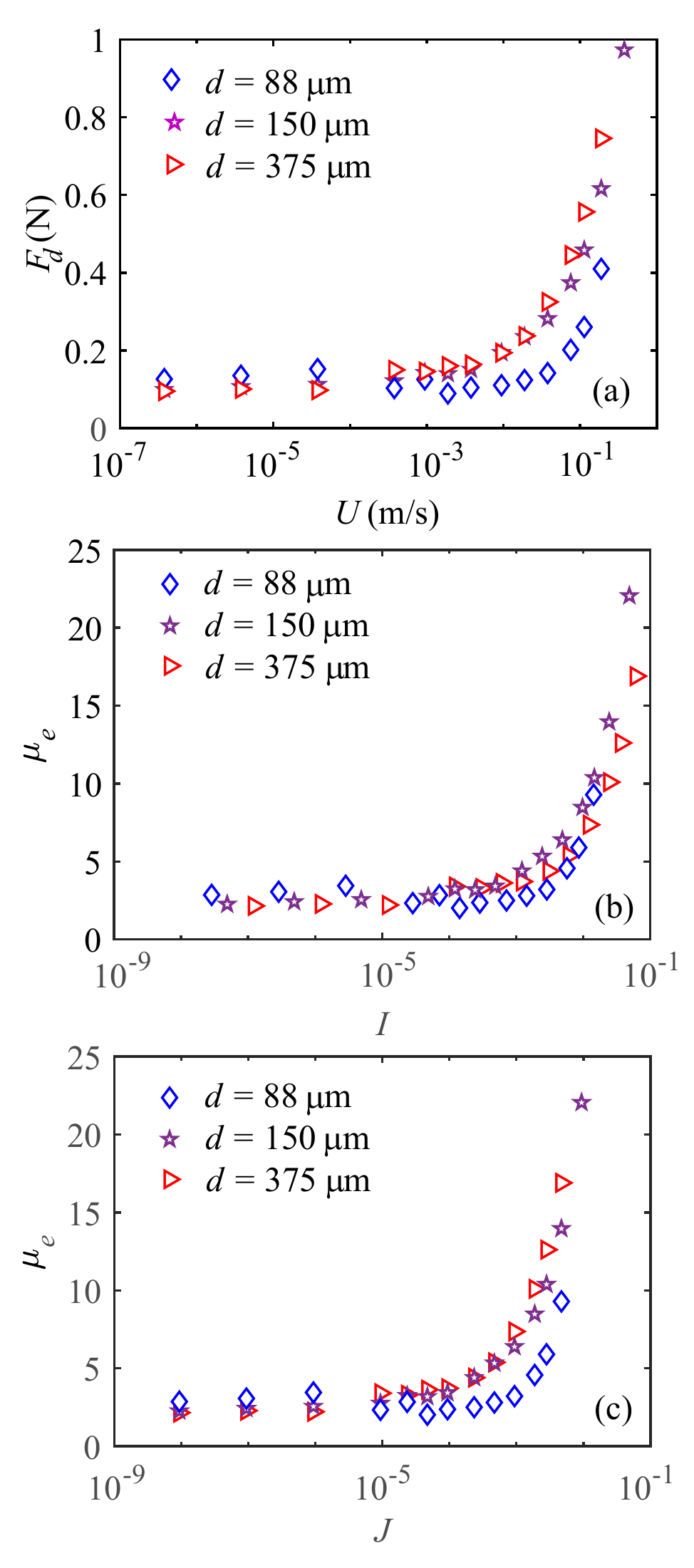}
\caption{(a) $F_d$ as a function of $U$ for different $d$ listed in Table~\ref{tab:grains} ($D=2.6$\,mm and $\eta_f=20$ mPa\,s). (b) Corresponding $\mu_e$ versus $I$. (c) Corresponding $\mu_e$ versus $J$.} 
\label{fig:diameter}
\end{figure}
To determine the influence of the particle size on the effective friction, we  examine the various grain sizes listed in Table~\ref{tab:grains} while using a rod with $D=2.6$\,mm which still satisfies the condition $d/D \ll 1$, and a  viscous fluid with $\eta_f=20$\,mPa\,s. The measured drag versus $U$, and corresponding $\mu_e$ versus $I$ and $J$ are plotted in Fig.~\ref{fig:diameter}. Because Eq.~\ref{eq:mue} and Eq.~\ref{eq:J} do not feature the grain size, the plots $F_d$ versus $U$ in Fig.~\ref{fig:diameter}(a) and $\mu_e$ versus $J$ in Fig.~\ref{fig:diameter}(c), look similar. At small $I$ and $J$ corresponding to low rod speeds, we observe that $\mu_e$ are similar in magnitude, and independent of the grain size.

However, we observe that $\mu_e$ does not fully collapse either in terms of $I$ or $J$ with increasing speed. This may be related to the fact that we have used the diameter of the rod as the scale in defining the shear rate $\dot{\gamma}$ in $I$ and $J$, when in fact other length scales, including $d$ and $z$ are also present in the system. $I$ is a better collapse due to a particle dependence in the definition, whereas $J$ has no dependence on particle size and thus Fig.~\ref{fig:diameter}(a) and Fig.~\ref{fig:diameter}(c) are essentially the same graph.  Thus, further work is required to identify an appropriate length scale in the rod system as speed is increased. 

\subsection{Rod diameter dependence}
\begin{figure}
\includegraphics[width=0.75\columnwidth]{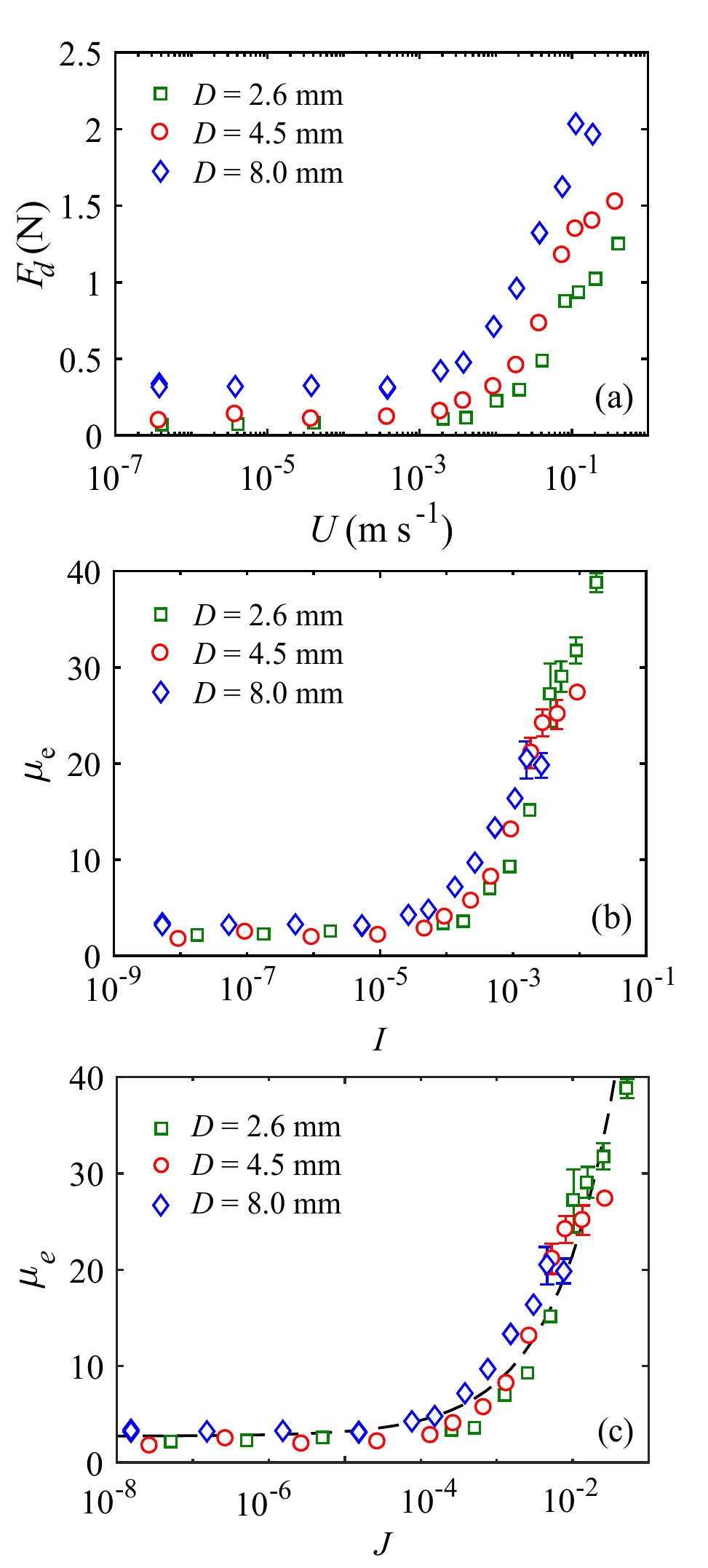}
\caption{(a) The drag as a function of $U$ for different rod diameter $D$ ($\eta_f = 100$\,mPa\,s). (b) $\mu_e$ as a function of $I$. (c) $\mu_e$ as a function of $J$. The curve given by Eq.~\ref{eq:fit} with the same fit shown in Fig.~\ref{fig:nondimensional}(c) is also plotted to guide the eye.} 
\label{fig:RodD}
\end{figure}

To test the effect of the diameter of the rod, we performed drag measurements with three different rod sizes $D=2.6$\,mm, $4.5$\,mm, and $8$\,mm corresponding to $D/d=13,23,40$ respectively.  These sizes were chosen so that $D/d \gg 1$, and $D/R \ll 1$ to limit unsteady motion and system size effects. Figure~\ref{fig:RodD}(a) shows $F_d$ as a function of $U$. In each case, $F_d$ is observed to increase with $U$, and $F_d$ is also observed to increase systematically with $D$. We plot the same data in terms of $\mu_e$ as a function of $I$ in Fig.~\ref{fig:RodD}(b) and Fig.~\ref{fig:RodD}(c). We observe a similar collapse of the data in terms of $I$ and $J$ because both are proportional to $U/D$ which are the two quantities being varied here. In the low speed limit, $\mu_e$ is observed to approach the same value $\mu_o$ irrespective of the size of the rod.  However, systematic variations are observed with increasing diameter, and the collapse of the data over the entire range of speeds is not particularly good.  

Further,  Eq.~\ref{eq:fit} is plotted in Fig.~\ref{fig:RodD}(c) with the same fitting constants $k$ and $n$ obtained in describing the data while varying viscosity alone in Fig.~\ref{fig:nondimensional}(c). We observe that the $\mu_e$ versus $J$ is observed to collapse on the same curve. Thus, we demonstrate that the form of $\mu_e$, and the constants $\mu_o$, $k$ and $n$ are independent of the rod diameter when $D \gg d$, as well as the fluid viscosity. 

\subsection{Depth dependence}
\begin{figure}
\includegraphics[width=0.95\columnwidth]{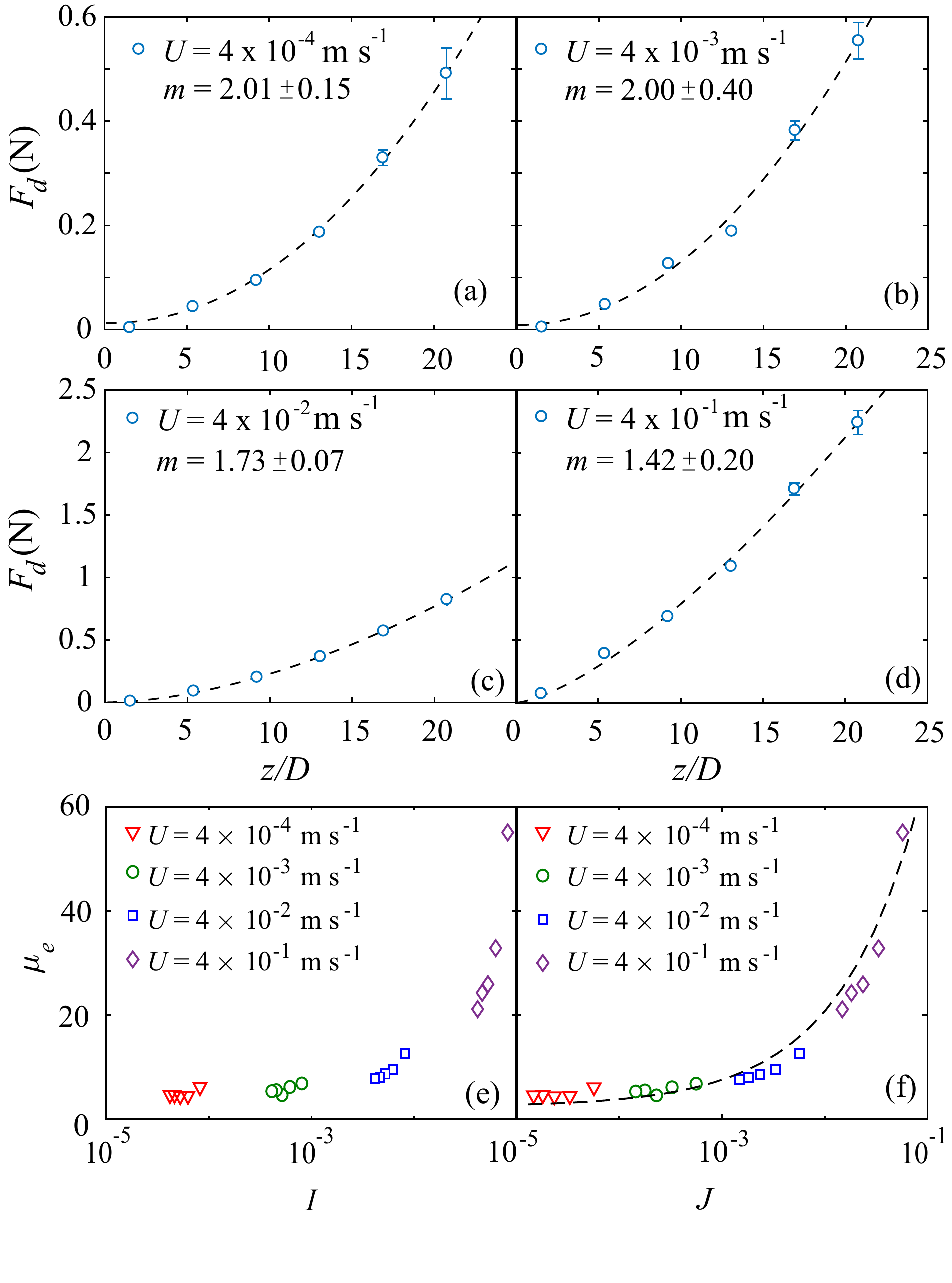}
\caption{ (a-d) Drag measured as a function of rod depth $z_r/D$ at different speeds ($\eta_f = 34$\,mPa\,s). The dashed line is a power law fit to the function $F(z) = F_o z^m$.  (e) $\mu_e$ versus $I$ corresponding to the data plotted in (a-d). (f) $\mu_e$ versus $J$ corresponding to the data plotted in (a-d). The curve given by Eq.~\ref{eq:fit} with the same fit shown in Fig.~\ref{fig:nondimensional}(c) is also plotted to guide the eye.} 
\label{fig:depth}
\end{figure}

We next examine the observed drag as a function of the depth $z$ to which the rod is inserted into the bed. Fig.~\ref{fig:depth}(a-d) shows the measured drag as a function of depth at various speeds. As $U$ increases, the increase in drag with depth is observed to become more linear. According to Eq.~\ref{eq:cyldragapprox}, the drag would increase approximately linearly with increased length inside the medium, with additional logarithmic corrections due to the denominator. But, according to Eq.~\ref{eq:schiffer}, the drag should be quadratic. Thus, we fit the data to a function $F(z) = F_o z^m$, with $F_o$ and $m$ as fitting constants to find the appropriate scaling with depth. At low $U$, we find $m \approx 2$ and thus recover the quadratic increase with depth described by Eq.\,\ref{eq:schiffer}. We then find that $m$ decreases to 1.4 over the range of $U$ measured. Thus, the observed response of the bed seems to transition from appearing granular-like to fluid-like, while not quite reaching the linear dependence which may be expected for a Newtonian fluid in the viscous regime. 

We analyze the effect of the rod depth on the drag in terms of $\mu_e$ versus $I$ in Fig.~\ref{fig:depth}(e), and $\mu_e$ versus $J$ in Fig.~\ref{fig:depth}(f). While both $\mu_e$ versus $I$ and $J$ increase with depth because both increase with $U$, a smoother dependence is observed with respect to $J$ where $J\propto P^{-1}$ and $I\propto P^{-1/2}$ . Further, we observe that corresponding $\mu_e$ increases with $J$ according to Eq.~\ref{eq:fit} with the same fitting constants $k$ and $m$. Thus, we observe that the effective friction encountered by a rod as its depth in the granular material is increased can be captured by $J$ at least in the case of sufficiently large $\eta_f$.

\section{Effective viscosity}

\begin{figure}
\includegraphics[width=0.75\columnwidth]{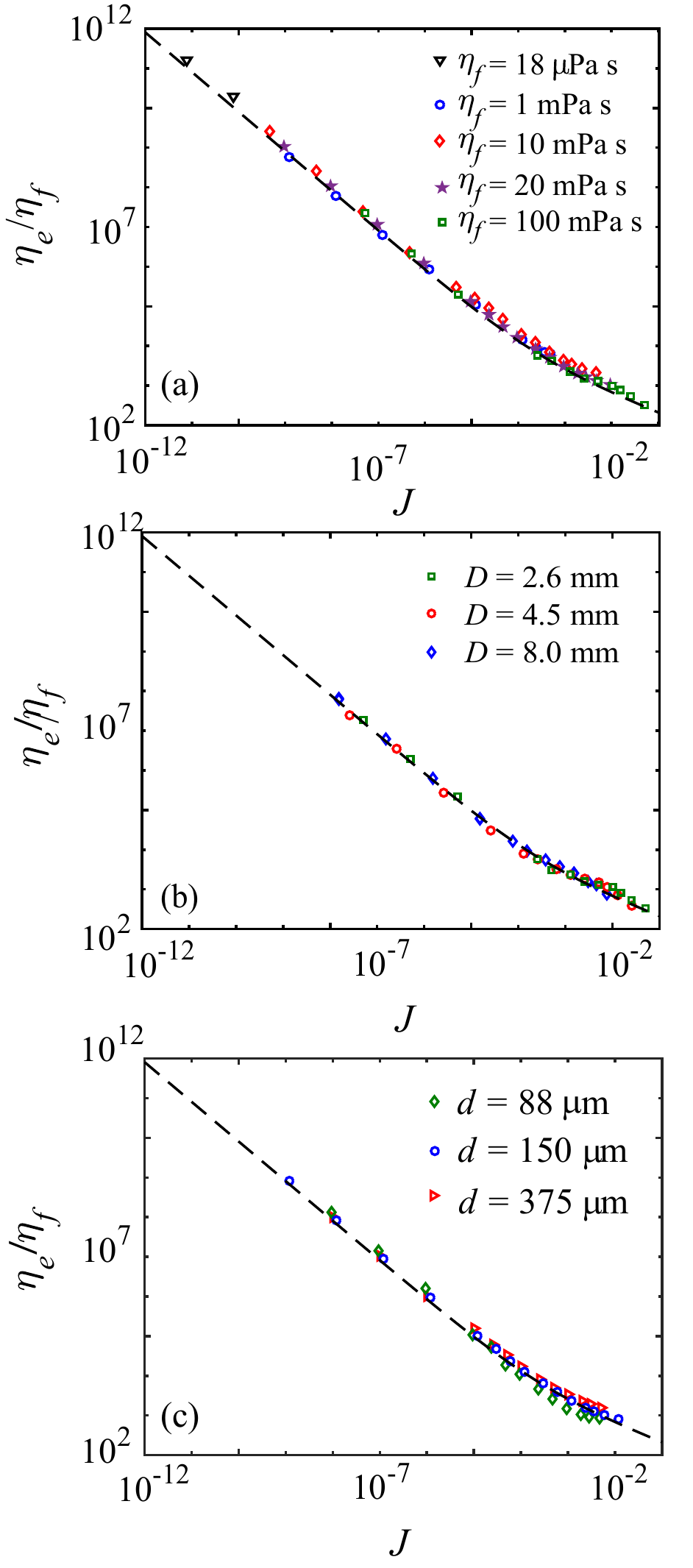}
\caption{(a) The effective viscosity $\eta_e$ versus $J$ for fluids with various $\eta_f$ collapses ($D = 2.6$\,mm).  The dashed line corresponds to Eq.~\ref{eq:etamue}. (b) $\eta_e$ versus $J$ for various $D$ is also observed to be described by the same curve. (c) $\eta_e/\eta_f$ as a function of $J$ for various $d$ roughly follows the same trend in all cases. However, small but systematic deviations can be also observed at the higher $J$. 
}\label{fig:etae}
\end{figure}

To analyze the observed drag from the perspective of rheology of the medium probed by the rod, one can define an effective viscosity by rearranging Eq.~\ref{eq:cyldragapprox}. We have, after substituting $z$ for $L$, 
\begin{equation}
\eta_e = \frac{F_d}{4 \pi z U} {[\frac{1}{2}-\log(\frac{z}{D})-\log(4)]}\,. 
\label{eq:etae}
\end{equation}
Now, $\eta_e$ can be related to $\mu_e$ by using Eq.~\ref{eq:etae} and Eq.~\ref{eq:J}. Then,   
\begin{equation}
\frac{\eta_e}{\eta_f} = \frac{\mu_e}{8 J} {[\frac{1}{2}-\log(\frac{z}{D})-\log(4)]}\,. 
\label{eq:etamue}
\end{equation}
Accordingly, we have plotted the measured $\eta_e/\eta_f$ versus $J$ for various  viscosities in Fig.~\ref{fig:etae}(a). Here, the functional form obtained after substituting Eq.~\ref{eq:mue} is also shown by the dashed line. 
Thus, in a regime where $\mu_e$ is constant, we can expect $\eta_e$ to essentially decrease inversely as $J$. But, as $\mu_e$ increases, $\eta_e$  can be expected to level-off. We observe that $\eta_f$ decreases with $J$ and collapses onto the dashed line over many orders of magnitude, both over the regime where $\eta_e$ appears to decrease inversely as $J$, and where it  decreases sublinearly. It should be noted here that the data even in the case of air is observed to collapse onto the same curve. 

Further, we have plotted $\eta_e/\eta_f$ versus $J$ measured by varying $D$ in Fig.~\ref{fig:etae}(b). The dashed line corresponding to $D = 2.6$\,mm is only drawn here because the lines corresponding to the other two measured $D$ essentially coincide because the terms in the numerator and denominator in terms of $D$ cancel out approximately over the range of $z/D$ studied. Here again we observe good collapse, in the case of all three data sets, on to the same curve. Finally, we have also plotted $\eta_e/\eta_f$ for various $d$ in Fig.~\ref{fig:etae}(c).  Except for some small but systematic deviations in the case of $d$ at large $U$ in Fig.~\ref{fig:etae}(c), the data is observed to again follow the same curve. 

\begin{figure}
\includegraphics[width=0.75\columnwidth]{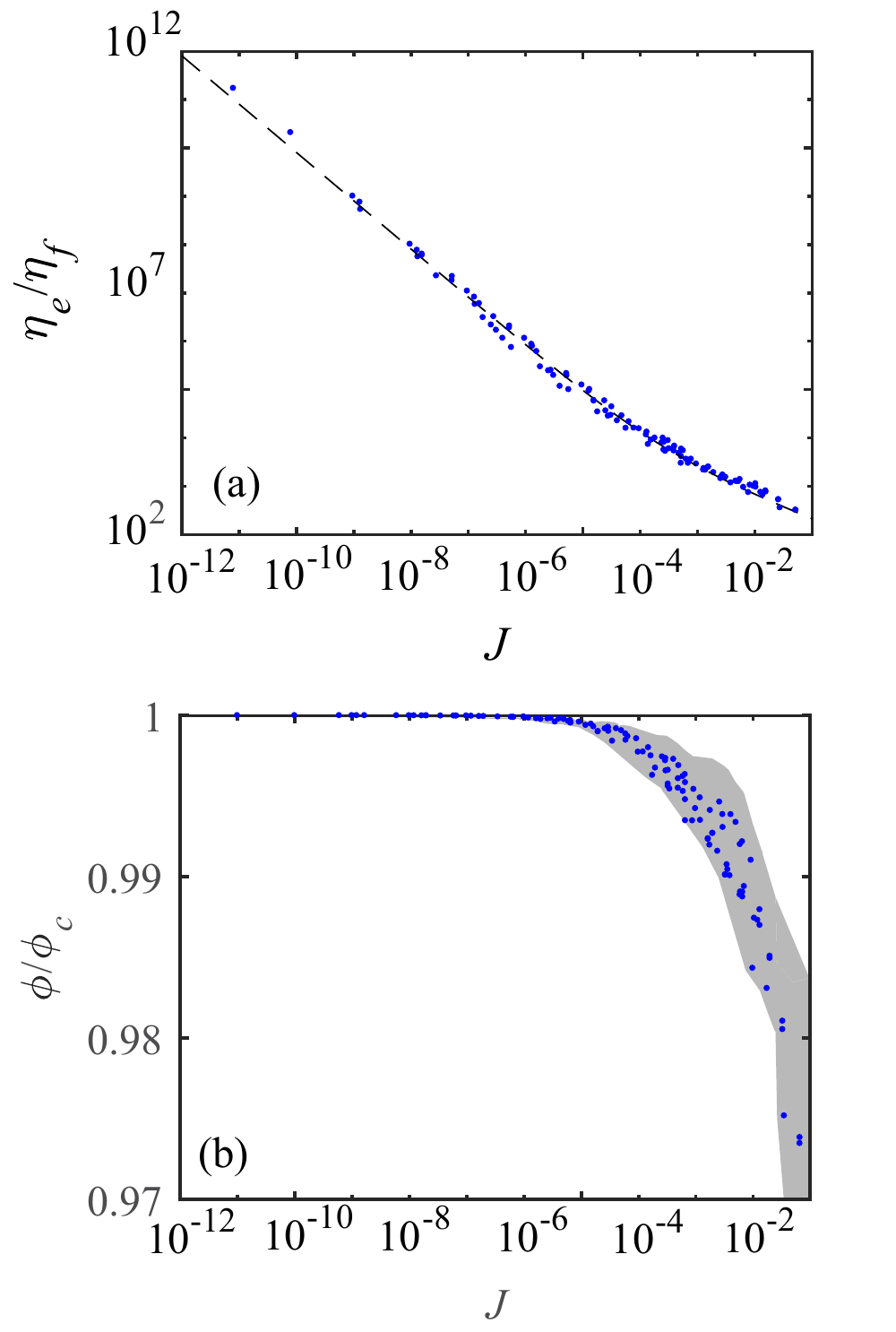}
\caption{(a) The effective drag $\eta_e$ versus $J$ collapses over the entire range of $z$, $\eta_f$, and $D$ investigated. This does not include $d$. (b) The packing fraction $\phi/\phi_c$ around the rod with the assumption of the Krieger-Dougherty relation shown in Eq.~\ref{eq:kriegerdougherty} between packing fraction $\phi$ and effective viscosity $\eta_e$. {\color{black} The gray shaded area indicates the error in the calculation of $\phi_c$ because of the uncertainty in calculating $\eta_e$ using Eq.~\ref{eq:etamue}, and  variation of the drag force $F_d$ with speed.}}\label{fig:Jall}
\end{figure}

To emphasize the good collapse of the data with many experimental parameters in terms of $J$, we have not only plotted all the data gathered by varying $\eta_f$ and $D$, but also $z$ in Fig.~\ref{fig:Jall}(a). The data is observed to be described by the line given by the same form as in Fig~\ref{fig:etae}.  Thus here, as in the case of $\mu_e$, we find that overall trends in $\eta_f$ can be captured by the single parameter $J$, while changing $U$, $D$, and $z$. The data corresponding to a single $d$ is shown because systematic deviations can be seen in Fig.~\ref{fig:etae}(c). The observed deviation with $d$ may be related to the fact that we have greatly simplified the estimate of $\dot{\gamma}$ when calculating $J$, and ignored other length scales such as $d$ and $z$ in the estimation of $\dot{\gamma}$.   

Now, examining the overall trend of $\mu_e$ with $J$, the medium can be interpreted to be shear thinning over the entire range of speeds. This is consistent with the physical picture that the advancing intruder shears the athermal granular bed which leads to dilation 
with speed. 
To arrive at an estimate of the corresponding decrease in volume fraction of the grains around the intruder, we use the Krieger-Dougherty empirical formula for the effective viscosity of a granular suspension as a function granular volume fraction $\phi$~\cite{krieger59},
\begin{equation}
\frac{\eta_e}{\eta_f} = \left(1 - \frac{\phi}{\phi_c}\right)^{-2.5\phi_c},
\label{eq:kriegerdougherty}
\end{equation}
where, $\phi_c$ is the critical volume fraction where the viscosity diverges.  Then, inverting this relation to obtain the implied variation of $\phi/\phi_c$, we plot the estimated $\phi/\phi_c$ versus $J$ from the measured $\eta_e/\eta_f$ in Fig.~\ref{fig:Jall}(b). The observed variation in $\phi/\phi_c$ is systematic and about 2\% over the range of speeds explored. Such small variations of packing fraction around the intruder are not possible for us to measure directly, but well within the variations observed in sedimented beds composed of frictional spherical grains~\cite{panaitescu12}. {\color{black} While the maximum random close packing volume fraction for spheres is $\approx 0.63$, typically a lower value $\approx 0.59$ is observed when the glass beads used in our experiments are sedimented to form a fresh bed [31]. When agitated, the packing fraction can compact by a few percent under very modest amount of applied shear. Thus, we can expect $\phi_c \approx 0.6$ after the bed is pre-sheared, and $\phi$ to decrease to 0.58 over the range of $J$ investigated based on Fig.~\ref{fig:Jall}(b).} 

\section{Conclusion}

In conclusion, we report a systematic study of the drag of a rod measured as a function of its speed, size and depth in a fluid saturated granular bed. The study is focused on conditions where viscous effects of the fluid can be important, and where the side walls can be considered to have negligible effect on the measured variations with experimental parameters. At low speeds, we recover the drag relations reported previously in Ref.~\cite{constantino11} where it was reported that the drag is constant and increases quadratically with depth.  As the speed is increased, we find that drag increases rapidly and to many-folds higher value than that found at vanishing speeds. The drag is then analyzed in terms of the effective friction which measures drag in relation to the average force acting on the rod due to the weight of the grains. 

We find that the observed variation of the effective friction with intruder speed, size, depth and fluid viscosity can be described to a large extent by the viscous number $J$. But some systematic variations are also observed especially while considering the grain size which points to a need to consider other length scales in the system besides the rod diameter considered here to obtain simple estimates. Further, we describe the observed effective friction as a function of $J$ in terms of an empirical formula which interpolates between the constant friction found at vanishing speeds and sub-linear increase found with increasing speeds. 

We also recast the measured drag in terms of the effective viscosity encountered by the rod to understand the response of the system from the perspective of a viscous fluid. Here, the effective viscosity probed by the rod is observed to be described essentially by the empirical effective friction function, the viscosity of the saturating fluid, and the viscous number. Finally, we show that only a small decrease of volume fraction of the granular component is needed to account for the significant shear thinning observed in the system.  

\section{Acknowledgments}
We thank Rausan Jewel for helping with experimental measurements, and National Science Foundation for support under Grant CBET-1805398. 

\bibliography{singlerod}

\appendix

\section{Interaction with side walls}
\label{sec:side}
\begin{figure}
\includegraphics[width=0.75\columnwidth]{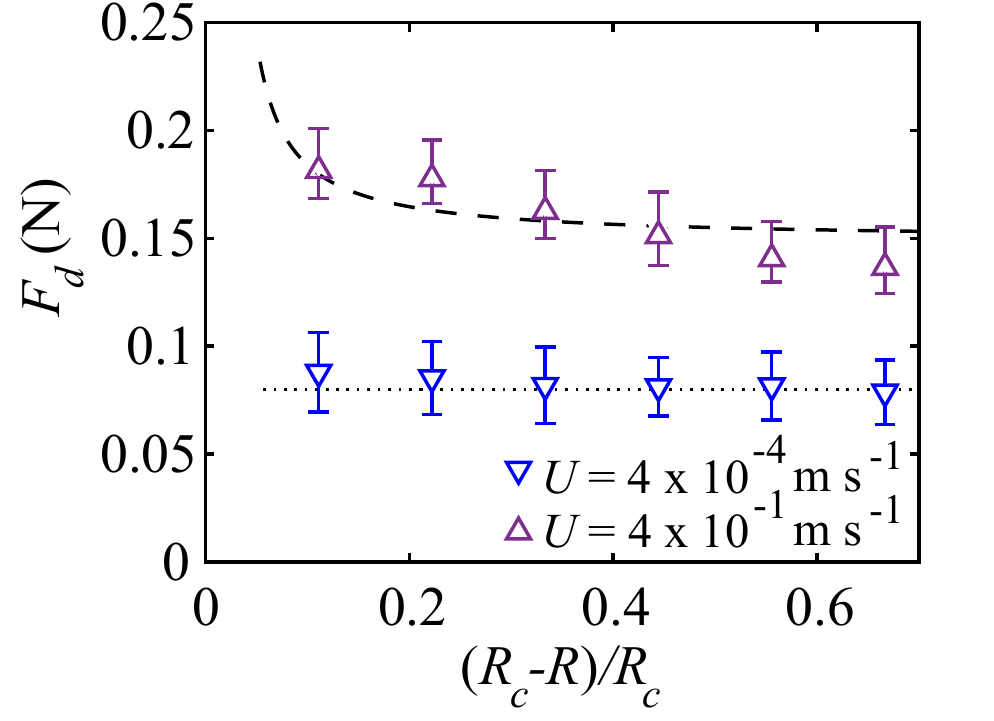}
\caption{The measured drag as a function of normalized distance from container boundary. The drag is essentially constant at low $U$ as described by the dotted line. At high $U$, drag increases near the side wall and is described by the dashed line  given by Eq.~\ref{eq:wallrod}. }
\label{fig:wall}
\end{figure}

We measure $F_d$ of the rod as a function of the radial distance $R$ from the container center to understand the effect of the finite size of the granular medium on the measured drag. Fig.~\ref{fig:wall} shows $F_d$ plotted as a function of $R$ scaled by $R_c$ corresponding to the low and high end of the range of speeds probed. In these experiments, the rod is inserted to depth $z_r = 3.5$\,cm, and fluid viscosity $\eta = 34$\,mPa\,s corresponding to the middle of the range of these parameters explored. We observe that $F_d$ is essentially constant at low speeds, but decreases with distance from the container sidewalls at higher speeds.

In the case of a granular bed, the stress near a wall can be affected by the frictional interaction of the grains~\cite{sperl06,guillard13} over depths greater than the distance to the walls. Because $z$ studied here is much smaller than the container size, the side walls can be expected to have negligible effect on measured drag as observed in Fig.~\ref{fig:wall}. There the measured drag is observed to be constant at the lowest speed.  

However, in the case of a rod moving in a viscous fluid parallel to a wall with non-slip boundary conditions, the drag at low Reynolds numbers can be written in terms of its length $z$ and the distance to the container side $(R_c - R)$ as~\cite{brennen77}:
\begin{equation}
F_d = \frac{4 \pi  z \eta_e U}{\log (\frac{2z}{D})+0.193-\frac{3 z}{2 (R_c-R)}}.
\label{eq:wallrod}
\end{equation}
This calculated form is shown in Fig.~\ref{fig:wall} using $\eta_e = 51.1 \eta_f$ as a fitting parameter. We observe that the slow increase in measured drag with decreasing distance to the side wall is roughly consistent with Eq.~\ref{eq:wallrod}. At still lower distance, a more rapid increase can be expected. However, since the goal of the study here is to examine the drag in an infinite sized bed, we have restricted our measurements to a region where the sidewall effect is small.

\end{document}